\documentstyle[12pt,thmsa,sw20aip]{article}
\input tcilatex
\begin{document}

\title{On local and global measurements of the speed of light on rotating platforms.}
\author{G. Rizzi and A. Tartaglia \\
%EndAName
Dip. Fisica, Politecnico di Torino, Italy; E-mail: tartaglia@polito.it\\
Received 14 October 1998; revised 19 March 1999}
\maketitle

\begin{abstract}
The paper shows that, conceptually and operationally, the speed of light as
measured locally in the inertial comoving frame of a point on the rim of a
rotating disk, is different from the one measured globally for a round trip
along the rim, obtained dividing the length of the rim (as measured in the
''relative space'' of the disk) by the time of flight of the light beam (as
measured by a clock at rest on the disk). As a consequence, contrary to some
recent claims, the anisotropy found in the global value, obtained by the
above procedure, in no way conflicts with the local isotropy, and the
internal consistency of the special relativity theory remains unchallenged.
\end{abstract}

\section{Introduction}

In 1997 Franco Selleri \cite{selleri}, in his long quest for inconsistencies
in the special relativity theory (SRT), pointed out a paradox concerning the
speed of light as measured on board a rotating disk. Actually his point
treats the speed of light along a closed circuit encircling the rotation
axis: when the platform is moving, the speed obtained dividing the length of
the contour by the time of flight, as measured in the ''relative space'' of
the disk (defined in Sec. 3) by an observer at rest on the platform, is
different whether measured in the rotation sense or in the opposite sense.
More explicitly, suppose a light beam is sent along the rim in the rotation
sense and another one in the opposite sense; then measure the average
velocities of both beams for a complete round trip, i.e. the ratio between
the length of the path and the times of flight read on a clock at rest on
the rim, and call them $c_{+}$ and $c_{-}$; then the ratio $\rho
=c_{+}/c_{-} $ differs from 1.

Since a rotating reference frame is not inertial, this anisotropy of the
light propagation is not, on itself, a surprising result; however Selleri
notices that when letting the platform's radius $R$ go to infinity and the
angular speed $\omega $ go to $0$ in such a way that the peripheral speed $%
\omega R$ of the turntable remains constant, the ratio $\rho $ too keeps a
constant value. However in the limit of infinite radius the uniform rotation
becomes a uniform translation, i.e. the local reference frame becomes
inertial. Here, according to SRT, the speed of light is assumed to be
exactly the same in any direction (since all inertial frames are assumed to
be optically isotropic); hence $\rho $ must be strictly $1$. This alleged
discontinuity in the behavior of $\rho $, under such limit process, is the
core of what we could call Selleri's paradox.

This issue has already been discussed elsewhere \cite{rita}, showing that a
full 4-dimensional relativistic treatment of the problem of the rotating
platforms avoids any discontinuity or inconsistency whatsoever, since the
speed of light, consistently defined, turns out to be exactly the same both
clockwise and counterclockwise, just as in an inertial reference frame.

However, though the 4-dimensional geometric point of view is clear and
consistent, nothing prevents from considering the problem from a different
viewpoint, rather natural for an observer living on the rotating disk. Then
some doubt is apparently allowed \cite{selleri}, \cite{selleri2}, since the
ratio $\rho $, when measured by means of meter rods and clocks (or rather a
single clock) at rest on the platform, actually differs from 1.

The root of Selleri's paradox can be identified in the basic assumption -
founded on the homogeneity of the disk along the rim - that the ''global
ratio'' $\rho =c_{+}/c_{-}$ of the average light velocities for complete
round trips, coincides with the ''local ratio'' $\rho _{o}$ of the forward
and backward light velocities. We shall however show, analyzing the actual
measurement procedures of the velocities in both cases, that $\rho $ cannot
in general be assumed to equal $\rho _{o}$, contrary to the claim by
Selleri. In fact, $\rho $ does not depend on the criterium for simultaneity
adopted along the rim, as Selleri correctly points out; $\rho _{o}$ does
instead strictly depend on the local simultaneity criterion. The two ratios,
which Selleri labels by the same letter $\rho $, refer to two different
kinds of measurements, so that there is no point in comparing them: they are
and remain different, whatever the size of $R$ is, be it finite or infinite,
with no harm for SRT. That this was the weak point of Selleri's argument has
been already remarked also by Budden \cite{budden}.

In sect. 2 the four-dimensional approach considered in ref. [2] is breafly
reexamined. Sect. 3 discusses two possible alternative definitions of space
of the platform along the rim, and compares the Minkowskian and the
operational approach to the interpretation of the measurements of space and
time intervals on board the rotating disk. Sect. 4 draws the general
conclusions.

\section{Constancy of the speed of light in Minkowski spacetime}

On a formal point of view, the SRT is the description of a four-dimensional
manifold, whose geometrical structure is uniquely determined by two
principles: the Einstein relativity principle and the principle of constancy
of the (one way) velocity of light in vacuum\footnote{%
Of course, the axiomatic basis of the SRT is not completely established by
the two principles mentioned above, but also embodies the so-called
''principle of locality'', which states the local equivalence of any
accelerated reference frame with a momentarily comoving inertial frame.}. As
well known, such manifold is the familiar Minkowski spacetime, in which the
time evolution of any massive particle is described in terms of a world line 
$\gamma _{m}$ which lies everywhere inside the light cone associated with
any point of $\gamma _{m}$.

This can be visualized in a standard spacetime diagram (in which space and
time are measured by the same unities and the coordinate lines are drawn
orthogonal to each other) as a world line whose slope, although variable, is
everywhere greater than 45$^{o}$. Only massless particles, particularly
photons, are described by null world lines, i.e. by world lines whose slope,
in the graphic representation, is always 45$^{o}$: any light beam in free
spacetime is described by a 45$^{o}$ slanting straight line, which can be
regarded as a generator of the light cone. This is a geometrical expression
of the principle of costancy of the one way velocity of light in minkowskian
spacetime. The interaction of the light beam with a mirror may change the
space direction of propagation of the beam, curving the trajectory in space
and the world line in spacetime, {\it without affecting its slope}. As a
consequence, when a light beam is lead to move along the rim of a rotating
disk, grazing a cylindrical mirror, its world line in $2+1$ dimensions turns
out to be a ''null helix'' wrapped around the world tube of the disk and
keeping everywhere a 45$^{o}$ slope.

A $2+1$ geometrical analysis of the Sagnac effect (see \cite{rita}) shows
how and why the times of flight for co-rotating and counter-rotating beams
are different, {\it although their world lines are helixes of constant (45}$%
^{o}${\it ) slope}; or, frasing it differently, {\it although their
velocities are the same - namely c - in any inertial frame, in particular in
the local inertial comoving frame at any point of the rim.}

To sum up, the special relativistic assumption of the constancy of the slope
of the world lines of light does not lead to inconsistencies or unphysical
discontinuities. On the operational point of view, this means that, in the
framework of SRT, the apparent global anisotropy of the propagation of light
along the rim is perfectly compatible with the local isotropy{\it , }%
contrary to Selleri's assumption.

\section{Actual measurements of the speed of light}

Once the internal consistency of the geometry of Minkowskian spacetime has
been established again, it still remains to confront it with the operational
procedures an observer at rest on the rotating disk uses, in order to
attribute actual values to the physical quantities of interest. The problem
is that any measurement concerning the geometry of the disk and the
synchronization of clocks on it is a well defined set of physical and
mathematical operations on an extended region of space (in particular along
the rim), whereas in a rotating frame special relativistic formulae are
merely local: any result obtained by extrapolating them globally cannot be
considered as a pure consequence of SRT, but depends on some (usually
hidden) further assumptions. In our opinion, the presence of recurrent
contradictions and paradoxes simply underlines the arbitrariness of such
extrapolations, from local to global.

Now, the obvious operational way to define and determine the (one way) speed
of a (massive or not) moving object is to measure the length of a given
travel and the time it takes, then divide the former by the latter. Of
course this procedure determines the slope of the world line of the moving
object only when the measurements are local (infinitesimal extension of the
space and time intervals); finite measurements can yield the slope only in
very special cases (constant slope world lines).

In the case of uniform rotation and of light travelling along the rim of the
rotating disk, the slope of the light world line is constant as well as that
of the observer's one; as a consequence, it can be determined not only by (a
sequence of) local measurements of space and time intervals, performed in
the local comoving frames, but also by global measurements of space and time
intervals referred to a complete (either co-rotating or counter-rotating)
round trip. However, in the second case the operational procedure should be
carefully defined, because of the presence of some unavoidable conventional
extrapolations, as pointed out before. More precisely: (i) the measure of
the length of a complete round trip depends on the definition of ''space on
the platform'', at least along the rim; (ii) the time taken by the light
beam for a complete round trip is an observable quantity (it is the proper
time lapse of a single clock), but the impossibility of a global
synchronization along the rim could require a suitable correction. The form
of the correction is imposed by the space-time geometry, according to the
particular definition chosen for the space of the platform (see later).

Among the many possible definitions of ''space on the platform along the
rim'', we consider in particular the following two: (i) the ''space of
locally Einstein simultaneous events'', defined as the set of events along
the rim such that any nearby pair of them are simultaneous according to the
Einstein criterium; (ii) the ''relative space'' $S:=T/I$, defined as the
quotient of the world tube $T$ of the disk by the congruence $I$ of the word
lines of the points of the disk.

The former is obtained extrapolating the local Einstein synchronization
procedure to the whole rim of the disk. This space coincides with the
space-like helix $\gamma _{S}$ considered at the beginning of Sec. 4 of ref.
2, and is everywhere Minkowski-orthogonal to the time-like helixes
corresponding to the world lines of the points of the rim.

The latter turns out to be the space of locations on the disk, regardless of
any kind of synchronization: ''two points of spacetime which lie on the same
disk word line ... are identified in the relative space'' \cite{tim}. This
space seems rather artificial on a Minkowskian point of view, but it should
appear quite natural for the observer on the platform, since the space
spanned by meter sticks arranged on the platform by this observer is
precisely the ''relative space'' of the disk.

Notice that we used both spaces, namely:\ the ''space of Einstein locally
simultaneous events'' when we adopt a Minkowskian approach, like in the main
part of \cite{rita}; and the ''relative space'' when we adopt an operational
approach, like in sect. 5 of \cite{rita} and everywhere in \cite{tartaglia}.

\subsection{{\bf \ }{\it Minkowskian approach}}

In this approach, the ''space of locally Einstein simultaneous events''
along the rim coincides with the space-like helix $\gamma _S$ considered
before; the important point is that $\gamma _S$ is not a circumference (this
is true only in the absence of rotation), but an open line whose slope
depends on the rotation velocity. Now, the proper length of an open line is
not a uniquely defined entity; we showed in particular in \cite{rita} that
the geometry of Minkowskian spacetime imposes different lengths for the
portion of $\gamma _S$ covered by the co-rotating and by the
counter-rotating light beams in a complete round trip. The difference in
these two lengths turns out to be

\begin{equation}
\delta s_{\gamma _{S}}=\frac{4\pi \left( \omega R\right) }{c\sqrt{1-\left(
\omega R\right) ^{2}/c^{2}}}R  \label{1}
\end{equation}
which exactly coincides, dividing by $c$, with the difference in time of
flight along the two round trips (see eq. (\ref{2}) later, which is
consistent with the Sagnac effect). As a consequence, this definition of
space ensures the equality of the global speed of light both for the
co-rotating and the counter-rotating light beams, restoring the isotropy of
light propagation.

We point out that this definition of space is the only one which can insure
the equality between global and local velocities, i.e. between the ''global
ratio'' $\rho $ and the ''local ratio'' $\rho _{o}$: this agrees with
Selleri's assumption, but both ratios equal exactly 1, with no harm for the
SRT.

\subsection{\it Operational approach}

In this approach, the ''relative space'' $S$ along the rim allows the
observer at rest on the platform to consider a unique length for the rim of
the disk (see sect. 5 of \cite{rita} and \cite{tartaglia}). The measure of
the two round trip times is obtained by one single clock (no need for
special synchronization procedures), and gives two different results. In
particular, the difference in time between the two round trips is

\begin{equation}
\delta \tau =\frac{4\pi \left( \omega R\right) }{c^{2}\sqrt{1-\left( \omega
R\right) ^{2}/c^{2}}}R  \label{2}
\end{equation}
which is an expression of the Sagnac effect \cite{sagnac}, \cite{stedman} .
In this case, the observer can draw the following conclusions, on the basis
of his measurements of space and time on the platform and without any
knowledge of Minkowskian spacetime structure (see \cite{rita}, sect.5): (i)
the platform on which he lives is rotating, and the desynchronization $%
\delta \tau $ of a pair of clocks, after slow round trips in opposite
directions, is a measure of the speed of this rotation; (ii) the durations
of travels along the closed path are not uniquely defined and, to obtain
reliable measures of them, the readings of clocks must be corrected by a
quantity $\pm \delta \tau /2$ to account for the desynchronization effect,
which is the same result obtained by Bergia and Guidone \cite{bergia}; (iii)
as a consequence of this correction, the speed of light is actually the same
both forward and backward.

On the other hand, if the readings are used without any theoretical
correction, the global measurement actually gives an anisotropic result at
all radii (as far as $\omega R\neq 0$); but this procedure, which is the one
proposed by Selleri, cannot prove his basic assumption, only founded on the
homogeneity of the disk along the rim, that the ''global ratio'' $\rho $
coincides with the ''local'' one $\rho _{o}$.

In fact, the measurements performed by the observer on the platform in order
to calculate the two ratios are completely different. The value of the
''global ratio'' turns out to be 
\begin{equation}
\rho =\frac{c_{+}}{c_{-}}=\frac{c-\omega R}{c+\omega R}\;\;  \label{3}
\end{equation}
and depends on the measurement of a difference of proper times read on a
single clock. This measurement is independent from any assumption about
synchronization.

On the contrary, the ''local ratio'' $\rho _{o}$ depends: (i) on the
measurement of two infinitesimal lengths (forward and backward) in the local
comoving frame; (ii) on the readings of three clocks (placed at the starting
point of the light beams and at the arrival points, in opposite directions),
Einstein synchronized in the local comoving frame. If Einstein
synchronization is used, the ''local ratio'' $\rho _{o}$ is exactly 1, and
cannot be identified with the ''global ratio'' $\rho $, which differs from 1.

One could object that also a local measurement of the light velocity can be
performed by means of a single clock, when the light beam is reflected by a
mirror placed at an infinitesimal distance from the source (two ways average
light speed). But in this case - that is what is usually made in actual
experiments, like e.g. Michelson-like experiments - the difference of
measurement procedures is still more evident: the global method, which
measures two one-way velocities of two light beams performing two complete
round trips along a closed path in opposite directions, cannot be used in
the local inertial frame, in which only the two ways light speed is
measurable. So the ''global ratio'' $\rho $ only is an observable; the
''local ratio'' $\rho _{o}$ is not. Selleri's assumption is 
\begin{equation}
\rho =\rho _{o}=\frac{c-\omega R}{c+\omega R}\neq 1\;\;\;\;\;\;%
%TCIMACRO{\TeXButton{per-ogni}{\forall } }
%BeginExpansion
\forall %
%EndExpansion
\omega \neq 0  \label{quattro}
\end{equation}
This assumption is equivalent to assuming a suitable non Einstein
synchronization in the local comoving frame, which could be called ''Selleri
synchronization'', consistent with the condition \cite{selleri2}, \cite
{budden}: 
\begin{equation}
c_{+}=c\left( 1+\frac{\omega R}{c}\right) ^{-1}\;;\;\;\;c_{-}=c\left( 1-%
\frac{\omega R}{c}\right) ^{-1}  \label{5}
\end{equation}
Such a synchronization requires of course a suitable non Lorentz coordinate
transformation, which in turn implies the existence of a privileged frame
and the absolute character of synchronization, see \cite{selleri}, \cite
{selleri2}, \cite{selleri3}.

An obvious consequence of eqs. (\ref{5}) is that light propagates
anisotropically in any local comoving frame along the rim, but the
observable two ways light speed is again $c$. As a consequence, the
''Selleri synchronization'' does not conflict with known experiments, but
conflicts with the standard Einstein synchronization (which assumes that
light propagates isotropically in any inertial frame: $c_{+}=c_{-}$ $=c$).

If Selleri's synchronization is used, the SRT is violated; in this case
Selleri's paradox only shows that, starting from an assumption violating the
SRT, a result violating the SRT follows.

We cannot treat, in the limits of this letter, the question of which
synchronization (Einstein or Selleri) is more adequate to the whole
experimental and theoretical context: we limit ourselves to claiming that
both are consistent and compatible with experiments, but the ''serious
logical problem in the SRT'' declared by Selleri does not exist.

\section{Conclusion}

To sum up our line of thought, we have seen that the direct measurement of
the speed of light along a closed path, free of any theoretical corrections,
does indeed reveal an anisotropy when the observer is rotating along the
contour. It would continue to be so also for a contour of infinitely great
curvature radius, though it is impossible to actually perform the experiment.

On the other side local measurements of the speed of light cannot evidence
any anisotropy. The global and local ratios between forward and backward
light velocities, which Selleri labels by the same letter $\rho $, refer to
two different kinds of measurements, which cannot be reduced one to the
other: they are and remain different, whatever the size of the platform
radius is, be it finite or infinite.The two classes of measurements do not
overlap and do not reveal, in the framework of the SRT, any internal
contradiction.

\end{document}